\documentclass[ALICE,manyauthors]{cernphprep}

\usepackage[english]{babel}
\usepackage{graphicx}
\usepackage{verbatim}
\usepackage{amsfonts}
\usepackage{makeidx}
\usepackage{color}
\usepackage{amsmath}
\usepackage{setspace}

\usepackage[colorlinks]{hyperref}
\usepackage[figure,table]{hypcap}

\hypersetup{
	bookmarksnumbered,
	pdfstartview={FitH},
	citecolor={blue},
	linkcolor={blue},
	urlcolor={black},
	pdfpagemode={UseOutlines}
}

\usepackage{epstopdf}
\usepackage{cite}
\usepackage{lineno}
\newcommand{\he}     {$^{3}\mathrm{He}$}
\newcommand{\antihe} {$^{3}\mathrm{\overline{He}}$}
\newcommand{\hyp}    {$^{3}_{\Lambda}\mathrm H$}
\newcommand{\antihyp}{$^{3}_{\bar{\Lambda}} \overline{\mathrm H}$}

\newcommand{\pip}    {$\pi^+$}
\newcommand{\pim}    {$\pi^-$}
\newcommand{\pio}    {$\pi$}
\newcommand{\dedx}   {d$E$/d$x$}
\newcommand{\PbPb}   {Pb--Pb}
\newcommand{\mom}    {\mbox{\rm MeV$\kern-0.15em /\kern-0.12em c$}}
\newcommand{\gmom}   {\mbox{\rm GeV$\kern-0.15em /\kern-0.12em c$}}
\newcommand{\mass}   {\mbox{\rm GeV$\kern-0.15em /\kern-0.12em c^2$}}
\newcommand{\Mmass}  {\mbox{\rm MeV$\kern-0.15em /\kern-0.12em c^2$}}
\newcommand{\pt}     {$p_{\rm T}$}
\newcommand{\ctau}   {$c \tau$}
\newcommand{\ct}     {$ct$}

\newcommand{\s}      {$\sqrt{s_{\mathrm{NN}}}$}
\newcommand{\dndy}   {d$N$/d$y$}

\begin{document}

%%%%%%%%%%%%%%%  Title page %%%%%%%%%%%%%%%%%%%%%%%%
%
\begin{titlepage}
\PHnumber{105}            
\PHdate{30 April}            
\PHyear{2015}
\title{\hyp~and \antihyp~production in \PbPb~collisions at \s~=~2.76 TeV}
\ShortTitle{\hyp~and \antihyp~production in \PbPb~collisions}   % appears on right page headers

%%% Do not change the next lines!
\Collaboration{ALICE Collaboration%
         \thanks{See Appendix~\ref{app:collab} for the list of collaboration
                      members}}
\ShortAuthor{ALICE Collaboration}      % appears on left page headers, do not change

\begin{abstract}

The production of the hypertriton nuclei \hyp\ and \antihyp\ has been measured for the first time 
in \mbox{\PbPb} collisions at \s~=~2.76~TeV with the ALICE experiment at LHC. The \pt-integrated  
\hyp\ yield in one unity of rapidity, 
\dndy\ $\times \mathrm{B.R.}_{\left( ^{3}_{\Lambda}\mathrm H \rightarrow ^{3}\mathrm{He},\pi^{-} \right)}$ = 
$\left( 3.86 \pm 0.77 (\mathrm{stat.}) \pm 0.68 (\mathrm{syst.})\right) \times$~10$^{-5}$ in the \mbox{0--10\%} 
most central collisions,  is consistent with the predictions from a statistical thermal model 
using the same temperature as for the light hadrons. The coalescence parameter $B_3$ shows a 
dependence on the transverse momentum,  similar to the $B_2$ of deuterons and the $B_3$ of \he\ nuclei.
The ratio of yields $S_3$ = \hyp/(\he $\times \Lambda/\mathrm{p}$) was measured to be 
\mbox{$S_3$ =  0.60 $\pm$ 0.13 (stat.) $\pm$ 0.21 (syst.)} in \mbox{0--10\%} centrality events; 
this value is compared to different theoretical models. 
The measured $S_3$ is compatible with thermal model predictions. 
The measured \hyp\ lifetime, $ \tau = 181^{+54}_{-39} (\mathrm{stat.}) \pm 33 (\mathrm{syst.})\ \mathrm{ps}$  
is in agreement within 1$\sigma$ with the world average value. 
\end{abstract}
\end{titlepage}

\setcounter{page}{2}

\section{Introduction and Physics Motivations}
High-energy heavy-ion collisions offer a unique way to study the behaviour of nuclear matter 
under conditions of extreme energy densities. 
At LHC energies, particles carrying strangeness are abundantly produced and light 
clusters of nucleons and hyperons, called hypernuclei, are expected to be formed \cite{Andronic:2011thermal}. 
Since their first observation \cite{Danysz:1974}, there has been a constant interest in searching 
for new hypernuclei as they offer an experimental way to study the hyperon-baryon ($YN$) 
and the hyperon-hyperon ($YY$) interactions, which are relevant for nuclear 
physics and nuclear astrophysics. For instance, the $YN$ interaction plays a key role 
in understanding the structure of neutron stars \cite{Weber:2006ph, Heiselberg:1999mq, Vidana:2013nxa, Lonardoni:2014hia}. 
The production of hypernuclei in heavy-ion collisions has been proposed and studied  for
a long time \cite{BraunMunzinger:1994iq, Armstrong:2002xh} 
and at ultrarelativistic energies it is possible to  produce particles otherwise inaccessible, 
such as anti-hypernuclei.
In fact, while many $\Lambda$-hypernuclei have been observed, the first observation of an 
anti-hypernucleus is rather recent and was reported from the analysis of Au--Au collisions at
$\sqrt{s_{\rm NN}}$ = 200 GeV by the STAR Collaboration at RHIC \cite{Abelev:2010sci}.
Since hypernuclei are weakly bound nuclear systems, they are sensitive probes  
of the final stages of the evolution of the fireball formed in the heavy-ion 
collisions \cite{Barrette:1994tw}. The yield of hypernuclei can distinguish between 
different production scenarios, usually described using two different theoretical approaches. 
The first one is based on a coalescence model \cite{Csernai:1986qf}, while the second one is 
based on the assumption that all the particle species can be described using a statistical 
thermal model \cite{BraunMunzinger:2003zd}. 
In the statistical thermal model a constant entropy over baryon ratio \cite{Siemens:1979dz} could explain why 
objects with such a small binding energy (few MeV) could survive the high temperature ($\approx$ 170~MeV) 
expanding fireball.
On the other hand, if hypernuclei are produced through coalescence of protons, neutrons and hyperons 
at freeze-out \cite{Gutbrod:1988gt}, they will provide a 
measurement of the local correlation between baryons and hyperons (strangeness)\cite{Steinheimer:2008hr}. \\
This letter presents a study of hypertriton and anti-hypertriton production at \s~=~2.76~TeV Pb--Pb 
collisions by the ALICE collaboration.The paper is organised as follows. In Section~\ref{sec:alice} the 
ALICE detector is briefly described. The data sample, analysis details and systematic uncertainties 
are presented in Section~\ref{sec:AnalysisDescription}. In Section~\ref{sec:Comparison} the obtained results 
are compared with theoretical models. Finally the conclusions are drawn in Section~\ref{sec:conclusions}.  

\section{The ALICE detector}
\label{sec:alice}
A detailed description of the ALICE detector can be found 
in~\cite{Abelev:2014ffa} and references therein.
For the present analysis the main sub-detectors used are the V0 detectors, 
the Inner Tracking System (ITS) and the Time Projection Chamber (TPC), which are 
located inside a 0.5~T solenoidal magnetic field.
The V0~\cite{Abbas:2013VZERO} detectors are placed around the beam-pipe on either side 
of the interaction point: one covering the pseudorapidity range $2.8 < \eta < 5.1$~\mbox{(V0-A)} 
and the other one covering $-3.7 < \eta < -1.7$~\mbox{(V0-C)}.
The collision centrality is estimated by using the multiplicity measured in the V0 detectors  
along with a Glauber model simulation to describe the multiplicity distribution as a 
function of the impact parameter \cite{Aamodt:2011oai,Abelev:2013qoq}.  
The ITS \cite{Aamodt:2010aa}  has six cylindrical layers of silicon detectors with radii between 3.9 and 
43~cm from the beam axis, covering the full azimuthal angle and the pseudorapidity range 
of $|\eta|<0.9$. The same pseudorapidity range is covered by the TPC \cite{Alme:2010TPC}, 
which is the main tracking detector. Hits in the ITS and found clusters in the TPC are used to 
reconstruct charged-particle tracks. These are used to determine the primary collision 
vertex with a resolution of about 10~$\mu$m in the direction transverse to the beams for heavy-ion collisions. 
The TPC is used for particle identification through the \dedx\ (specific energy loss) in the TPC gas.
\section{Analysis}
\label{sec:AnalysisDescription}
\begin{figure}[!htbp]
\begin{tabular}{ccc}
\begin{minipage}{.5\textwidth}
\centerline{\includegraphics[width=1\textwidth]{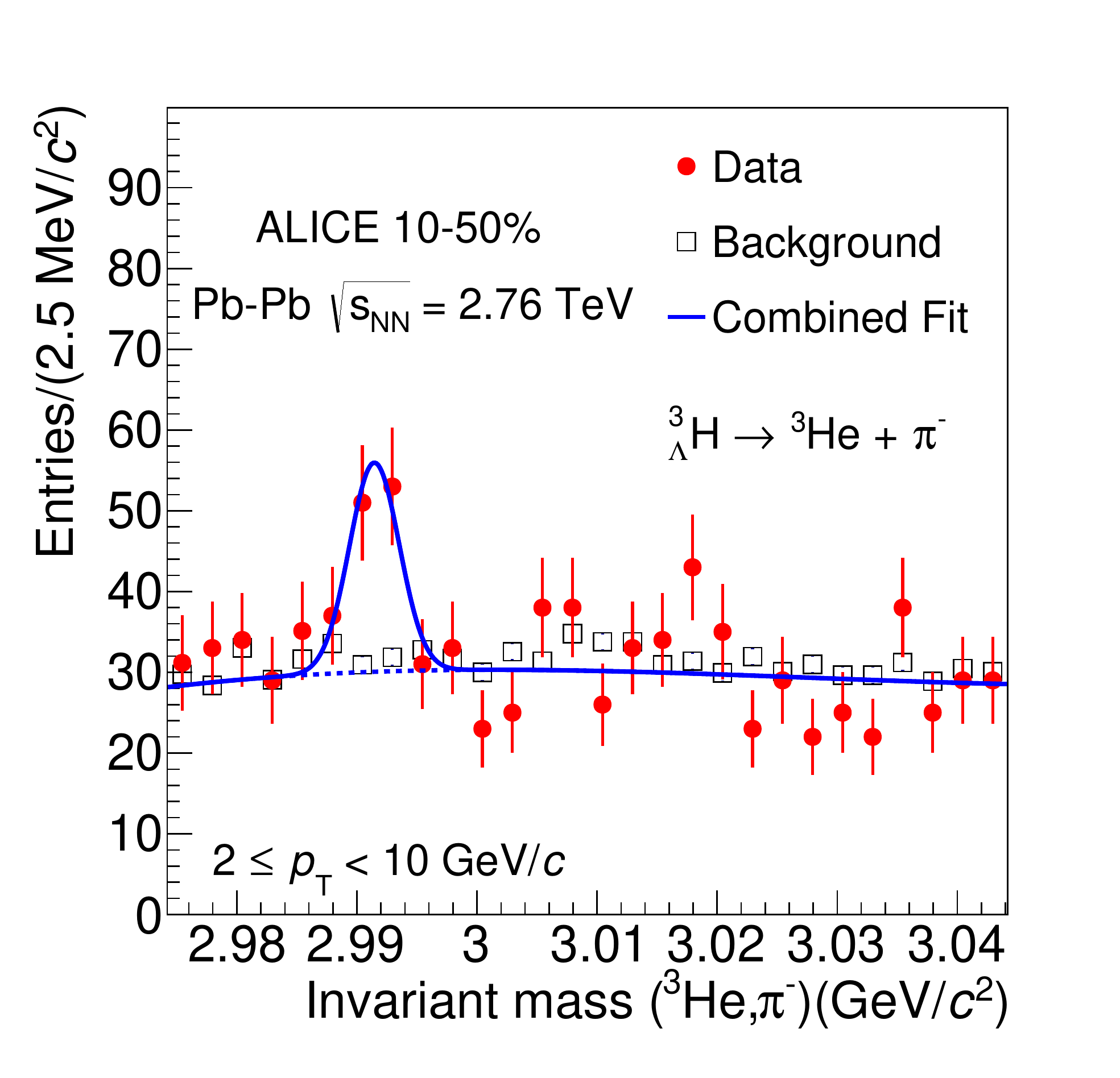}}
\end{minipage} & 
\begin{minipage}{.5\textwidth}
\centerline{\includegraphics[width=1\textwidth]{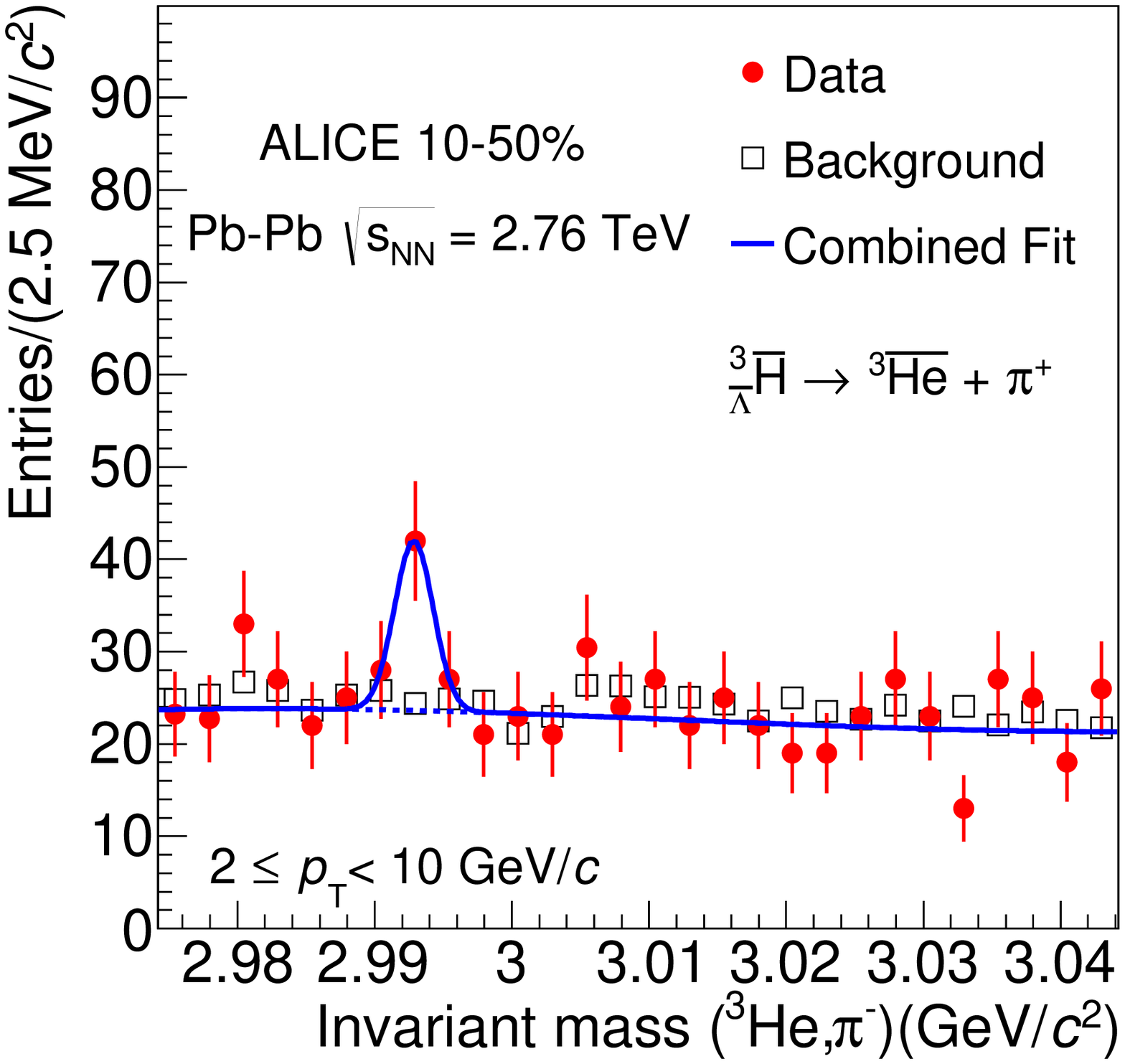}}
\end{minipage} 
\end{tabular}
\caption{Invariant mass of (\he,\pim) (left) and (\antihe,\pip) (right) for events 
with 10--50\% centrality in the pair 2~$\leq$~\pt~$<$~10~\gmom\ interval. 
The data points are shown as filled circles, while the squares represent the background distribution 
as described in the text. The curve represents the 
function used to perform the fit and used to evaluate the background and the raw signal. 
The significance in $\pm$3$\sigma$ around the peak is 3.5 and 3.0 for the invariant mass 
distribution of (\he,\pim) and (\antihe,\pip), respectively.}
\label{fig:InvMass}
\end{figure}

The (anti-)hypertriton (\antihyp) \hyp\ is the lightest observed 
hypernucleus and is a bound state formed by a (anti-)proton, a (anti-)neutron 
and a (anti-)$\Lambda$. 
The \hyp\ and \antihyp\ production yields were measured by detecting their mesonic decay 
(\hyp~$\rightarrow$~\he~+~\pim) and (\antihyp~$\rightarrow$~\antihe~+~\pip)  
via the topological identification of secondary vertices and the analysis of the invariant 
mass distributions of (\he +\pim) and  (\antihe + \pip) pairs.\\
The analysis was done using \PbPb\ collisions at \s~=~2.76~TeV taken in 2011. 
The events were collected with an interaction trigger requiring a signal in both V0-A and V0-C. 
Only events with a primary vertex reconstructed within $\pm$10~cm, along the beam axis, 
from the nominal position of the interaction point were selected.
The analysed sample, collected with two different centrality trigger configurations corresponding 
to the \mbox{0--10\%} and \mbox{10--50\%} centrality intervals, contained approximately 
\mbox{20$\times$10$^6$} and \mbox{17$\times$10$^6$} events, respectively. \\
The \hyp\ can be identified via the invariant mass of its decay products and, since it 
has a lifetime similar to the free $\Lambda$ (\ctau\ $\sim$ 8 cm), 
in most cases it is possible to identify its decay up to a few cm away from the 
primary vertex. The decay vertex was determined by exploiting a set 
of geometrical selections: i) the distance of closest approach (DCA) 
between the two particle tracks identified using \dedx\ in the TPC as \he\ and \pio, 
ii) the DCA of the \pio$^{\pm}$~tracks from the primary vertex, 
iii) the cosine of the angle between the total momentum of the decay pairs at the secondary vertex 
and a vector connecting the primary vertex and the secondary vertex 
(pointing angle), and iv) a selection on the proper lifetime (\ctau) of the candidate.
An additional selection on the \hyp\ (\antihyp) rapidity ($|y|<$0.5) was applied. \\
Figure~\ref{fig:InvMass} shows the invariant mass distribution of (\he,\pim) on the left and 
(\antihe,\pip) on the right for events with \mbox{10--50\%} centrality 
in the pair transverse momentum range 2 $\leq$ \pt\ $<10$ \gmom. 
In order to estimate the background, for each event the \pio\ track detected at the secondary vertex
was rotated 20 times by a random azimuthal angle. The shape of the corresponding (\he, \pio) invariant mass 
distribution was found to reproduce the observed background outside the signal region. 
The data points were fitted with a function which is the sum of a Gaussian and a third
degree polynomial, used to describe the signal and the background, respectively. 
The background was normalized to the measured values in the \mbox{3.01 -- 3.08 \mass} region.
The fit to the background distribution was used to fix the parameters of the polynomial in 
the combined fit. 

\begin{figure}[!htbp]
\begin{tabular}{ccc}
\begin{minipage}{.5\textwidth}
\centerline{\includegraphics[width=1\textwidth]{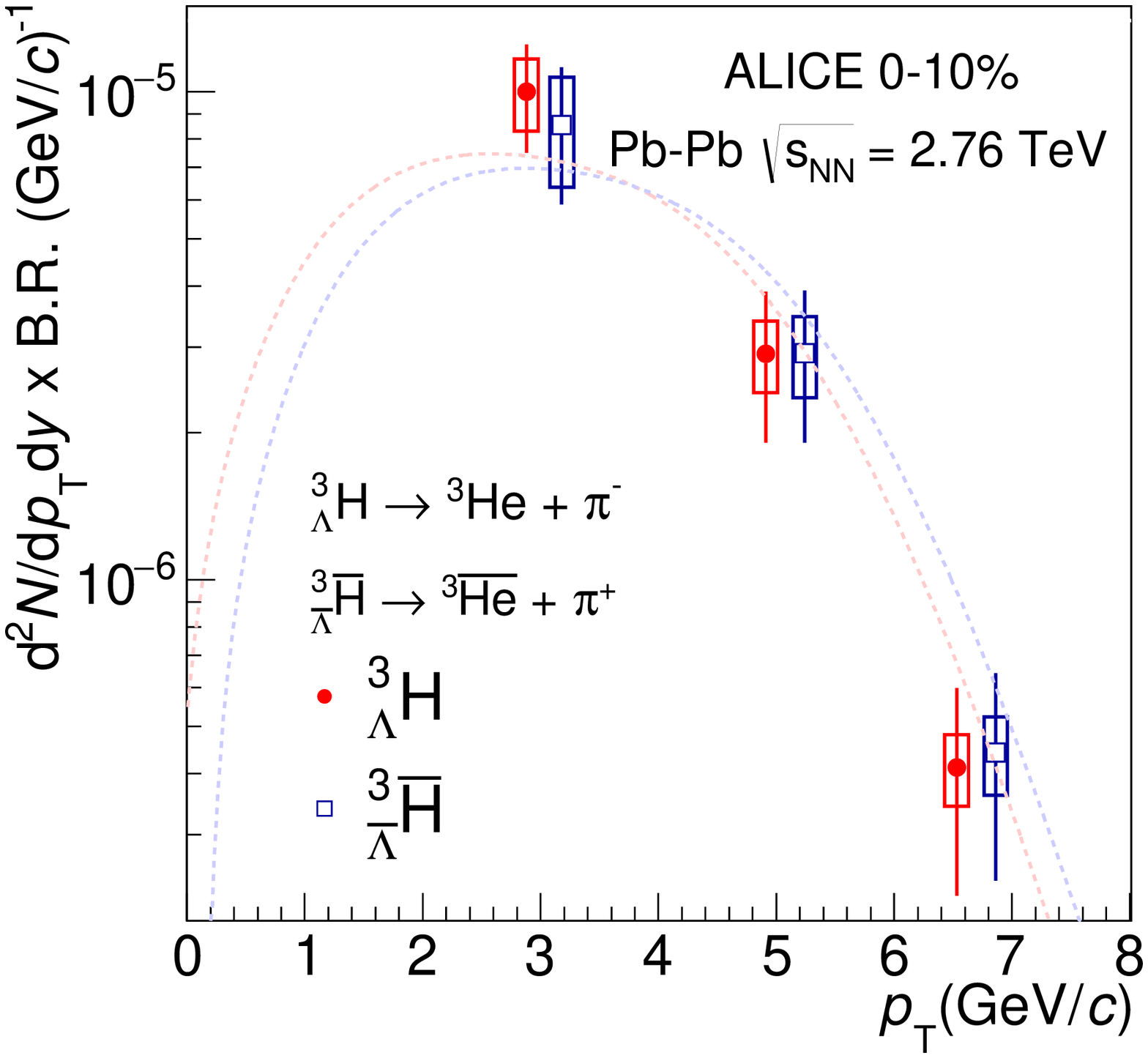}}
\end{minipage} & 
\begin{minipage}{.5\textwidth}
\centerline{\includegraphics[width=1\textwidth]{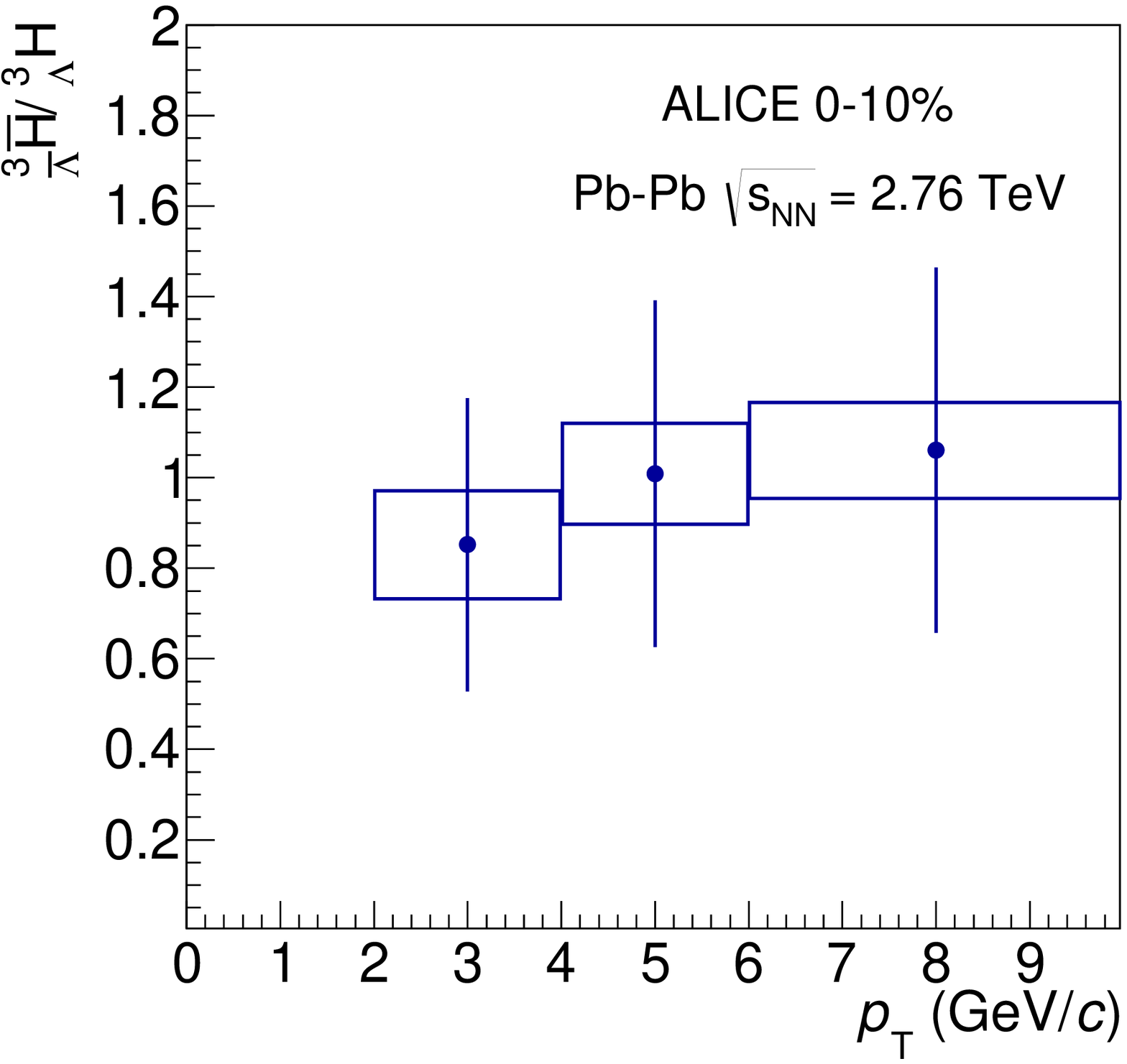}}
\end{minipage} 
\end{tabular}
\caption{Left: Transverse momentum spectra multiplied by the B.R. of the 
\hyp~$\rightarrow$~\he~+~\pim\ decay for \hyp\ (filled circles) 
and \antihyp\ (squares) for the most central (0--10\%) \PbPb\ collisions 
at \s\ = 2.76~TeV for $|y| <$ 0.5. Symbols are displaced for better visibility. 
The dashed lines are the Blast-Wave curves used to extract the particle yields 
integrated over the full \pt\ range. In order to take into account the large binning 
used in the analysis and the limited number of bins, the center of each bin was evaluated 
weighting the actual bin center with the Blast-Wave function.  
Right: \antihyp\ to \hyp\ ratio as a function of \pt. 
In both panels statistical uncertainties are represented by bars and systematic 
uncertainties are represented by open boxes.}
\label{fig:Spectra}
\end{figure}

In the 0--10\% most central collisions, a signal was extracted in three transverse momentum 
intervals (\mbox{2 $\leq$ \pt\ $<$ 4 \gmom}, \mbox{4 $\leq$ \pt\ $<$ 6 \gmom}, 
\mbox{6 $\leq$ \pt\ $<$ 10 \gmom}) , for both \hyp\ and \antihyp. 
In the 10--50\% centrality class a signal both for \hyp\ and \antihyp\ was obtained for the full \pt\ range under study 
(2~$\leq$~\pt~$<$~10~\gmom).
From the combined fit results the mean value, the width and the yield of the signal 
were extracted. 
The mean invariant mass ($\mu$ = 2.991 $\pm$ 0.001 (stat.) $\pm$ 0.003 (syst.)~\mass) is 
compatible within uncertainties with the mass from the literature \cite{Juric:1973mh}.
The signal width, $\sigma$~=~(3.01~$\pm$~0.24~(stat.))$\times$10$^{-3}$\mass\, obtained as the   
mean value of all the measured widths, is reproduced by Monte Carlo simulations 
and is driven by detector resolution. 
The raw yield of the signal was defined as the integral of the Gaussian function in 
a $\pm$~3~$\sigma$ region around the mean value. 
The significance of both matter and anti-matter signals varies in the 
different \pt\ bins in the range of 3.0--3.2 $\sigma$ for the most central collisions (0--10\%) 
and ranges from 3 to 3.5 $\sigma$ for the semi-central ones (10--50\%).\\
A correction factor which takes into account the detector acceptance, the reconstruction 
efficiency, and the absorption of \hyp\ (\antihyp) by the material crossed was determined 
as a function of \pt. 
Detector acceptance and reconstruction efficiency were evaluated using a dedicated 
HIJING Monte Carlo simulation~\cite{Wang:1991h}, where the only allowed decay was 
the two-body decay to charged particles, (\hyp~$\rightarrow$~\he~+~\pim) and (\antihyp~$\rightarrow$~\antihe~+~\pip). 
The simulated particles were propagated through the detector using the GEANT3 transport 
code~\cite{Brun:1994} and then processed with the same reconstruction chain as for the data.\\
Since the absorption of (anti-)(hyper)nuclei is not properly implemented in GEANT3, 
a correction based on the p ($\overline{\mathrm{p}}$) absorption was applied in order to 
take into account the absorption of \hyp\ (\antihyp) and \he\ (\antihe) by the material 
of the ALICE detector.  
In this approach, the \he\ and \hyp\ were treated as states of three independent   
p ($\overline{\mathrm{p}}$). 
The \he\ was considered as a bound state of 3 protons because 
the proton absorption correction in the ALICE detector was measured \cite{Abbas:2013rua}.  
The direct measurement offers the advantage of having a probability density which takes into
account the effective material of the detector crossed by a
charged particle. The effect of using protons instead of neutrons was tested with deuterons, 
which were considered as a bound state of 2 protons and the absorption correction 
was evaluated with the same model used for \he. 
The result was compared with the one obtained with the absorption correction of GEANT3 patched 
with hadronic cross sections for d and $\overline{\mathrm{d}}$. 
The two calculated absorption corrections where found to be consistent within uncertainties. 
To take into account the small $\Lambda$ separation energy   
($B_{\Lambda}({^{3}_{\Lambda}\mathrm{H}}) = 0.13 \pm 0.05$~MeV~\cite{Bando:1990yi}), 
the absorption cross section of the \hyp\ was increased by 50\% with 
respect to the one of the \he.
This choice was based on the theoretical calculation of \hyp\ absorption cross-section 
\cite{Evlanov:1998py} on $^{238}$U and its ratio with the extrapolation of 
\he\ cross section on the same target \cite{Kox:1985ex}. Using the same extrapolation it was possible to 
evaluate the same ratio on ALICE materials.
The correction applied to the extracted yield was about 12\% for \hyp\ 
and about 22\% for \antihyp.
The total systematic uncertainty takes into account, as lower and upper limits 
of the \hyp(\antihyp) absorption cross section, values respectively equal to or two times 
higher than the absorption cross section of \he(\antihe). 
This uncertainty is \pt\ dependent, and its values are reported in Table~\ref{table:summsystPt}. 
Other sources of systematic uncertainties in the yield evaluation were estimated:
\begin{itemize}
\item The systematic uncertainty due to the single-track efficiency, and the different choices 
of the track quality selections was taken from~\cite{Abelev:2013ra}. 
A 10\% uncertainty is quoted for the two body decay of \hyp.
\item \hyp\ lifetime: since the \hyp\ lifetime is not accurately known, 
the influence of varying the \hyp\ lifetime 
on the efficiency was evaluated by variation of the proper lifetime of the injected \hyp\ 
in the Monte Carlo simulation. 
The associated uncertainty was estimated using two additional dedicated 
Monte Carlo simulations with different 
lifetimes. The injected lifetime of \hyp\ (\antihyp) was varied ($\pm$1$\sigma$) with 
respect to the result obtained in this analysis, leading to an uncertainty of 8.5\%. 
\item The uncertainty related to the signal extraction procedure was evaluated by constraining 
 fit parameters ($\mu$ and $\sigma$) in different ways. This source led to a 9\% uncertainty.
\end{itemize}
The systematic uncertainty due to the uncertainty of the ALICE detector material budget 
and \pt\ distribution in the Monte Carlo used for the efficiency estimation led to a 1\% systematic uncertainty.\\
The \hyp\ and \antihyp\ spectra are shown in Figure~\ref{fig:Spectra}~(left panel), 
multiplied by the branching ratio (B.R.) of the 
\hyp~$\rightarrow$~\he~+~\pim\ decay. The anti-hypertriton  to hypertriton ratio as a function of \pt\ is 
shown in Figure~\ref{fig:Spectra}~(right panel). It is consistent with unity over the whole 
considered \pt\ range, as expected from zero net baryon density at LHC energies.  
In the ratio, the common systematic uncertainties (tracking efficiency, lifetime, and signal 
extraction method) cancel out and have therefore been removed.

In order to take into account the unmeasured \pt\ region and to extract the particle yields 
integrated over the full \pt\ range, the spectra were fitted using a blast-wave 
function \cite{Schnedermann:1993} whose parameter values were taken 
from the deuteron analysis \cite{Abbas:2014nuclei} 
leaving the normalization free. The function 
fits the data with a $\chi^2$/NDF of 0.92. %The function is shown as a dashed line in Figure~\ref{fig:Spectra}. 
The extrapolation in the \pt$<$ 2~\gmom\ region contributes 28\% to the final yield for both 
\hyp\ and \antihyp, while the contribution for \pt\ $>$ 10 \gmom\ is negligible. 
Different transverse momentum distributions were used to evaluate the systematic uncertainty related 
to the extrapolation, which was found to be 5\%. 

\begin{table}[!h]
\def\arraystretch{1.1}% 
\begin{center}
\begin{tabular}{ l | c c c c | c c c c}
\hline 
\hline                             &\multicolumn{4}{c|}{\hyp}                    & \multicolumn{4}{c}{\antihyp} \\ 
\hline                             &\multicolumn{4}{c|}{\pt~intervals (\gmom)}   &\multicolumn{4}{c}{\pt~intervals (\gmom)}\\ 
                                   & 2--4   & 4--6   & 6--10 & F.R.    & 2--4    & 4--6   & 6--10   & F.R.\\
\hline  Absorption                 & 5.4\%  & 5.3 \% & 5.4\%  & 5.4\%  & 13\%    & 10\%   & 8.9 \%  & 10.6\%\\  
        Tracking efficiency        & 10 \%  & 10 \%  & 10 \%  & 10 \%  & 10 \%   & 10 \%  & 10 \%   & 10 \% \\
        \hyp\ lifetime             & 8.5\%  & 8.5\%  & 8.5\%  & 8.5 \% & 8.5\%   & 8.5\%  & 8.5\%   & 8.5 \%\\
	    Signal extraction method   & 9  \%  & 9 \%   & 9 \%   & 9 \%   & 9 \%    & 9 \%   & 9 \%    & 9 \% \\
        Extrapolation at low \pt\  &   -    &  -     &  -     & 5 \%   & -       &  -     &   -     & 5 \% \\
		Total                      & 16.8\% & 16.8\% & 16.8\% & 17.5\% & 20.5\%  & 18.8\% &  18.2\% & 19.8  \% \\
\hline
\end{tabular} 
\caption {Summary of systematic uncertainties for the three \pt\ intervals and in the full range (F.R.) considered. 
These uncertainties are the same for events with 0--10\% and 10--50\% centrality. 
For the final systematic uncertainty evaluation they were added in quadrature.}
\label{table:summsystPt}
\end{center}
\end{table}

To determine the lifetime, the (\hyp\ + \antihyp) sample was divided into four intervals in 
\ct~=~${MLc}/{p}$, where $M$ is the mass, $L$ the decay length, $c$ is the speed of light, 
and $p$ is the total momentum. The mass was fixed to the value from the literature  
$M$ = 2.991~\mass\ \cite{Juric:1973mh}. 
For the determination of the lifetime, both centrality classes \mbox{0--10\%} and 
\mbox{10--50\%} were used. The signal was extracted in the intervals: 
\mbox{1 $\leq$ \ct\ $<$4~cm}, \mbox{4 $\leq$ \ct\ $<$ 7~cm}, \mbox{7 $\leq$ \ct\ $<$ 10~cm} 
and \mbox{10 $\leq$ \ct\ $<$ 28~cm}. 
To estimate the lifetime, the raw signal was corrected by the detector acceptance, 
the reconstruction efficiency and the absorption of \hyp\ (\antihyp) in the material. 
The same dedicated HIJING Monte Carlo simulation and the same procedure used to determine the \pt\ dependence 
of the efficiency were used. The sources of systematic uncertainty are shown in Table~\ref{table:summsystct}.

\begin{table}[!h]
\begin{center}
\begin{tabular}{l c }
\hline
\hline Source & Value \\
\hline Signal extraction method  & 9\%  \\
       Tracking efficiency       & 10\% \\
       Absorption                & 12\% \\ 
       Total                     & 18\% \\
\hline
\end{tabular} 
\caption {Summary of systematic uncertainties for the determination of the proper lifetime of \hyp+\antihyp.}
\label{table:summsystct}
\end{center}
\end{table}

\begin{figure}[!ht]
\begin{center}
\includegraphics[width=0.5\textwidth]{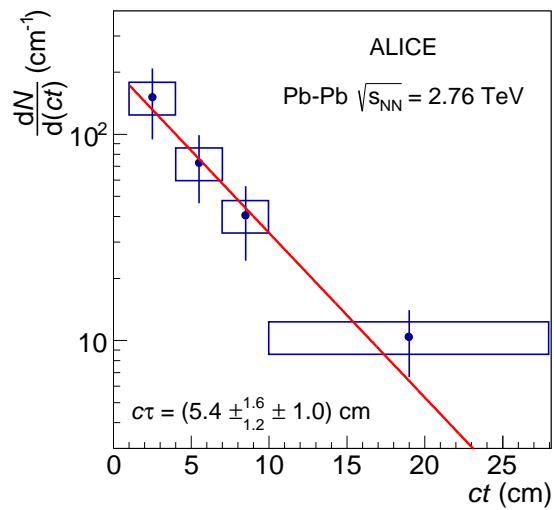}
\end{center} 
\caption{Measured d$N$/d(\ct) distribution and an exponential 
fit used to determine the lifetime. The bars and boxes are the statistical and 
systematic uncertainties, respectively.}
\label{fig:fitctau}
\end{figure}

An exponential fit was performed  to determine the lifetime. The d$N$/d(\ct) distribution and the exponential 
fit are shown in Figure~\ref{fig:fitctau}. The vertical bars show the statistical uncertainties and the 
boxes represent the systematic uncertainties. The slope of the fit results in a proper decay length 
of \ctau\ = $\left(5.4^{+1.6}_{-1.2}(\mathrm{stat.})\pm 1.0 (\mathrm{syst.})\right)$~cm.

\begin{figure}[!ht]
\begin{center}
\includegraphics[width=1\textwidth]{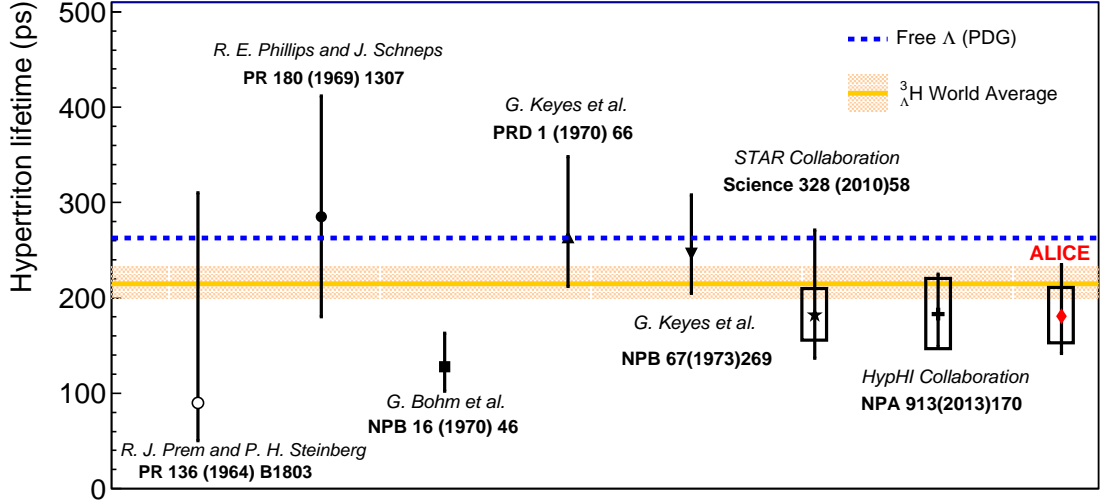}
\end{center} 
\caption{\hyp\ lifetime ($\tau$) measured by in this analysis (red diamond) compared 
with published results. The band represents the world average of 
\hyp\ lifetime measurements $\left(\tau = 215^{+18}_{-16}\right)$ ps, 
while the dashed line represent the lifetime of $\Lambda$ as reported by the Particle Data Group \cite{Agashe:2014kda}.}
\label{fig:lifetimeworldaverage}
\end{figure}
The lifetimes of light $\Lambda$-hypernuclei ($A\leq 4$) are expected 
to be very similar to that of the free $\Lambda$, if the $\Lambda$ in the hypernucleus is weakly bound \cite{Kamada:1997rv}. 
The measured lifetimes of light hypernuclei such as 
\hyp~\cite{Prem:1964hyp, Keyes:1968zz, Phillips:1969uy, Bohm:1970se,Keyes:1970ck, Keyes:1974ev,
Rappold:2013fic, Abelev:2010sci} are not known as precisely as the $\Lambda$ lifetime, and 
theoretical predictions \cite{Dalitz:1959zz, Dalitz:1962eb, Rayet:1966fe, Ram:1971tf, 
Mansour:1979xw, Kolesnikov:1988uy,Congleton:1992kk, Gloeckle:1998ty, Kamada:1997rv} 
are scattered over a large range, too.
Recently, a statistical combination of the experimental 
lifetime estimations of \hyp\ available in literature was published, resulting in an 
average value $\tau = \left(216^{+19}_{-18}\right)\ \mathrm{ps}$~\cite{Rappold:2014jqa}.\\
With the present data, a lifetime of $\tau = \left(181^{+54}_{-39} (\mathrm{stat.})\pm 33 (\mathrm{syst.})\right)$ ps 
has been obtained. It is compared with the previously published results in Figure~\ref{fig:lifetimeworldaverage}.
Our result, together with the previous ones, was used to re-evaluate the world average of the existing results 
using the same procedure as described in \cite{Rappold:2014jqa}. The obtained value,  
$\tau = \left(215^{+18}_{-16}\ \mathrm{ps}\right)$, is shown as a band in 
Figure~\ref{fig:lifetimeworldaverage}. The result obtained in this analysis is compatible 
with the computed average.  

\section{Comparison between experimental yields and theoretical models}
\label{sec:Comparison}
The product of the \pt-integrated yield and the B.R. of the \hyp~$\rightarrow$~(\he~+~\pim) 
decay for \hyp\ and \antihyp\ for two centrality classes (0--10\% and 10--50\%) 
are reported in Table~\ref{table:yields}. The systematic uncertainties also include  
the contribution due to the low \pt\ extrapolation as 
described in Section \ref{sec:AnalysisDescription}. \\
It is possible to compare the \pt-integrated  \hyp\ yield at different centralities 
by scaling them according to the charged-particle densities 
$\langle \mathrm{d} N_{ch}/\rm d\eta \rangle$. 
For central (0--10\%) collisions $\langle \mathrm{d} N_{ch}/\rm d\eta \rangle$~=~1447 $\pm$ 39, 
while for semi-central (10--50\%) $\langle \mathrm{d} N_{ch}/\rm d\eta \rangle$~=~575 $\pm$ 12.
The ratio 
\begin{equation}
\frac{\left(\frac{\left( ^{3}_{\Lambda}\mathrm H + ^{3}_{\bar{\Lambda}} \overline{\mathrm H} \right)_{(0-10\%)}} {\left( ^{3}_{\Lambda}\mathrm H + ^{3}_{\bar{\Lambda}} \overline{\mathrm H} \right)_{(10-50\%)}}\right)} {\left(\frac{\langle \mathrm{d} N_{ch}/\rm d\eta \rangle _{(0-10\%)}}{\langle \mathrm{d} N_{ch}/\rm d\eta \rangle _{(10-50\%)}}\right)} = 1.34 \pm 0.35 (\mathrm{stat.}) \pm 0.24 (\mathrm{syst.}) 
\end{equation}
is compatible with unity within 1~$\sigma$. The \hyp\ (\antihyp) production scales with centrality like the 
charged-particle production.

\begin{table}[!h]
\def\arraystretch{1.5}% 
\begin{center}
\begin{tabular}{ c | c | c | c }
\hline
\hline 
       Centrality  & $\langle \mathrm{d} N_{ch}/\rm d\eta \rangle$  & \hyp\ \dndy\ $\times\ \mathrm{B.R.} \times 10^{5} $  &   \antihyp\ \dndy\ $\times\ \mathrm{B.R.} \times 10^{5} $ \\
\hline  0--10\%    &             1447 $\pm$ 39                      & $3.86 \pm 0.77(\mathrm{stat.}) \pm 0.68(\mathrm{syst.}) $   &        $3.47 \pm 0.81(\mathrm{stat.}) \pm 0.69(\mathrm{syst.}) $  \\   
        10--50\%   &             575  $\pm$ 12                      & $1.31 \pm 0.37(\mathrm{stat.}) \pm 0.23(\mathrm{syst.}) $   &        $0.85 \pm 0.29(\mathrm{stat.}) \pm 0.17(\mathrm{syst.}) $  \\   
\hline 
\end{tabular} 
\caption {\pt-integrated \hyp\ yield times the B.R. of the \hyp~$\rightarrow$~(\he~+~\pim) decay, 
for \hyp\ and \antihyp\ in \PbPb\ collisions at \s~=~2.76~TeV for different 
centrality classes in $|\rm{y}|<0.5$. For each centrality interval the average 
$\langle \mathrm{d} N_{ch}/\rm d\eta \rangle$ is also 
reported \cite{Aamodt:2011oai}.}
\label{table:yields}
\end{center}
\end{table}
\subsection{Comparison between thermal models and experimental yields} 
\begin{figure}[!ht]
\begin{center}
\includegraphics[width=0.5\textwidth]{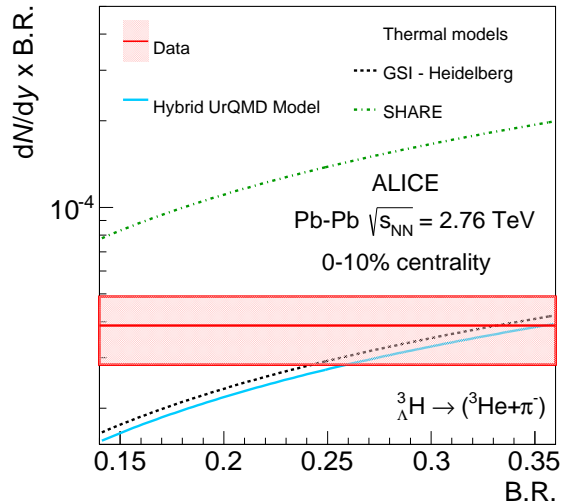}
\end{center} 
\caption{\pt-integrated  \hyp\ yield times branching ratio as a function of branching 
ratio (\dndy\ $\times \mathrm{B.R.}$ vs B.R.). 
The horizontal line is the measured value and the band represents statistical and systematic 
uncertainties added in quadrature. Lines are different theoretical expectations as 
explained in the text.}
\label{fig:dndyvsBR}
\end{figure}

Since the decay branching ratio of the \hyp~$\rightarrow$~\he~+~\pim\ was estimated only  
relative to the charged-pion channels \cite{Keyes:1974ev}, the corresponding value (B.R. = 35\%) 
provides an upper limit for the absolute branching ratio. 
On the other hand, a theoretical estimation for the \hyp~$\rightarrow$~\he~+~\pim\ decay 
branching ratio, which also takes into account decays with neutral mesons decays, gave a B.R. = 25\% \cite{Kamada:1997rv}. 
Assuming a possible variation on the B.R. in the range 15--35\%, we show in Figure~\ref{fig:dndyvsBR} 
a comparison of our result with different theoretical model calculations 
\cite{Andronic:2011thermal, petran:2013therm, uqrmd:missing}. The measured \dndy~$\times~\mathrm{B.R.}$ 
is shown as a horizontal line, where the band represent statistical and systematic 
uncertainties added in quadrature while the different theoretical models are shown as lines. 
The data are compared with the following models: two versions of the statistical 
hadronization model \cite{Andronic:2011thermal,petran:2013therm} and the hybrid UrQMD model 
\cite{uqrmd:missing}, which combines the hadronic transport approach with an 
initial hydrodynamical stage for the hot 
and dense phase of a heavy-ion collision. 
The two versions of the statistical hadronization model used are the equilibrium statistical 
model (GSI-Heidelberg), described in \cite{Andronic:2011thermal} and references therein, 
with a temperature $T_{\mathrm{ch}}$~=~156~MeV and the non-equilibrium thermal model (SHARE), 
described in \cite{petran:2013therm} and references therein, 
with $T_{\mathrm{ch}}$~=~138.3~MeV, $\gamma_q$ = 1.63 and $\gamma_s$ = 2.08, where 
$\gamma_q$ and $\gamma_s$ represent the quark and strangeness phase space occupancy 
of the system created after the collision, respectively.\\
The non-equilibrium thermal model (SHARE) \cite{petran:2013therm}
overestimates the (anti-)hypertriton \pt-integrated yield by a factor from 
2 to 5 depending on the branching ratio (B.R.).  
For the branching ratio expected following \cite{Kamada:1997rv} (B.R. = 25\%) the equilibrium thermal model 
\cite{Andronic:2011thermal} (GSI-Heidelberg) and the hybrid UrQMD model \cite{uqrmd:missing} describe 
the data best.

\begin{figure}[!ht]
\begin{center}
\includegraphics[width=0.5\textwidth]{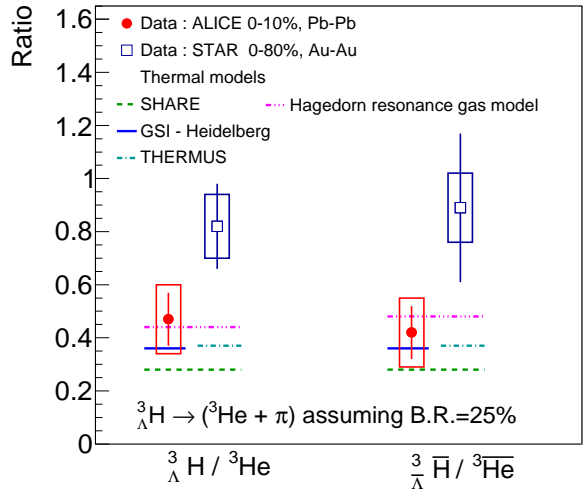}
\end{center} 
\caption{The ratios \hyp/\he\ and \antihyp/\antihe\ determined by the present analysis 
(filled circles) for matter and anti-matter compared with STAR results 
(squares) \cite{Abelev:2010sci} and theoretical predictions (lines) 
\cite{Andronic:2011thermal, petran:2013therm, Cleymans:2011pe, pal:2013hyp} 
as described in the legend.}
\label{fig:ratiohyphe}
\end{figure}

A fit, based on the thermal fit described in \cite{Andronic:2011thermal}, was performed to the hypertriton 
yield and to yields from other light flavour hadrons, except K$^*$, previously measured by our Collaboration 
at \s~=~2.76~TeV~\cite{Abelev:2013vea, ABELEV:2013zaa, Abelev:2013lk0, Abelev:2014uua, Abbas:2014nuclei}. 
The inclusion of the deuteron, \he\ \cite{Abbas:2014nuclei} and \hyp\ in the thermal fit \cite{Floris:2014pta} in 
addition to lighter particles, does not change the resulting freeze-out temperature 
(\mbox{$T_{\mathrm{ch}}$ = 156 $\pm$ 2 MeV}) and the measured yields of the nuclei and the hypertriton agree 
with the model predictions within 1~$\sigma$.
The results on the hypertriton yields discussed above were also used to determine the \hyp/\he\ 
and \antihyp/\antihe\ ratios, which are shown in Table.~\ref{table:ratiosHypHe}. 
In order to compute the ratios, our previous measurement of \he\ and \antihe\ yields~\cite{Abbas:2014nuclei} 
were used. These results were compared with different theoretical 
models \cite{petran:2013therm, Cleymans:2011pe, pal:2013hyp} and results from the STAR 
experiment~\cite{Abelev:2010sci} at \s~=~200~GeV, which use the same B.R.~=~25\%. 
The comparison is shown in Figure~\ref{fig:ratiohyphe}. 
STAR results are higher than ALICE results, but still compatible within uncertainties.

\begin{table}[!h]
\def\arraystretch{1.5} 
\begin{center}
\begin{tabular}{ c | c | c }
\hline
\hline        Centrality         &    \hyp\ / \he\                           &   \antihyp\ / \antihe\  \\
\hline          0--10\%         &   $0.47 \pm 0.10 (\mathrm{stat.}) \pm 0.13(\mathrm{syst.})$ &   $0.42 \pm 0.10 (\mathrm{stat.}) \pm 0.13(\mathrm{syst.})$ \\
               10--50\%         &    $0.40 \pm 0.11 (\mathrm{stat.}) \pm 0.11(\mathrm{syst.})$&   $0.26 \pm 0.09(\mathrm{stat.}) \pm 0.08(\mathrm{syst.})$  \\
\hline 
\end{tabular} 
\caption {Ratios of \hyp/\he\ and \antihyp/\antihe\ assuming a B.R. = 25\% 
for the \hyp~$\rightarrow$~\he~+~\pio\ decay~\cite{Kamada:1997rv}. 
The results from \he\ and \antihe\ analysis measured by the ALICE experiment were used \cite{Abbas:2014nuclei}.}
\label{table:ratiosHypHe}
\end{center}
\end{table}

\subsection{Data comparison to coalescence models and $S_3$ ratio}
At the moment no prediction of the \hyp\ and \antihyp\ yields in a non-trivial 
dynamical coalescence model is available at LHC energies.
Nevertheless within a simple coalescence model it is possible to evaluate some parameters 
which are sensitive to the existence of coalescence mechanisms for hypernuclei formation.
In the empirical coalescence model \cite{Csernai:1986qf} the cross section for the production 
of a cluster with mass number $A$ is related to the probability that $A$ nucleons have relative 
momenta less than $p_0$, which is a free parameter of the model.  
This provides the following relation between the production cross sections of the nuclear 
cluster emitted with a momentum $p_A$ and the nucleon emitted with a momentum $p_{\mathrm{p}}$
\begin{equation}
E_A \frac{\mathrm{d}^3N_A}{\mathrm{d}^3p_A} = B_A \left(E_{\mathrm{p}} \frac{\mathrm{d}^3N_{\mathrm{p}}}{\mathrm{d}^3p_{\mathrm{p}}}\right)^A,
\label{eq:coal}
\end{equation}
where $p_A = Ap_{\mathrm{p}}$. For a given nucleus, the coalescence parameter $B_A$  
should not depend on the momentum since it depends only on the cluster parameters:
\begin{equation}
B_A =  \left( \frac{4 \pi}{3} p_0^3 \right) ^{(A-1)} \frac{M}{m^A}
\end{equation}
where $M$ and $m$ are the nucleus and the proton mass, respectively and $p_0$ is the relative momentum 
between the constituent nucleons of the nucleus.  
The parameter $B_3$ was computed for \hyp\ according to Equation \ref{eq:coal} using the spectrum 
 shown in Figure~\ref{fig:Spectra} and our previous measurement of the proton~\cite{Abelev:2013vea} and 
$\Lambda$~\cite{Abelev:2013lk0} spectra.\\
Parameters $B_2^{\mathrm{d}}$ and $B_3^{^{3}{\rm{He}}}$ obtained in \cite{Abbas:2014nuclei} are compared with the 
hypertriton  $B_3^{^{3}_{\Lambda}\mathrm{H}}$ from this analysis using the relations
\begin{equation}
B_2^{^{3}\mathrm{He}} = \sqrt{\frac{m_{\mathrm{d}}^2}{m_{^3{\mathrm{He}}} m_{\mathrm{p}}} B_3^{^{3}\mathrm{He}}},
\label{eq:eq4}
\end{equation}

\begin{equation}
B_3^{^{3}_{\Lambda}\mathrm{H}} = B_3^{^{3}\mathrm{He}}\frac{m_{\mathrm{p}} m_{^{3}_{\Lambda}\mathrm{H}}}{m_{^3{\mathrm{He}}}m_{\Lambda}}.
\label{eq:eq5}
\end{equation}

and finally 
\begin{equation}
B_2^{^{3}_{\Lambda}\mathrm{H}} = \sqrt{\frac{m_{\mathrm{d}}^2 m_\Lambda}{m_{\mathrm{p}}^2 m_{^{3}_{\Lambda}\mathrm{H}}} B_3^{^{3}_{\Lambda}\mathrm{H}}}.
\label{eq:eq6}
\end{equation}

\begin{figure}[!htbp]
\begin{tabular}{ccc}
\begin{minipage}{.5\textwidth}
\centerline{\includegraphics[width=1\textwidth]{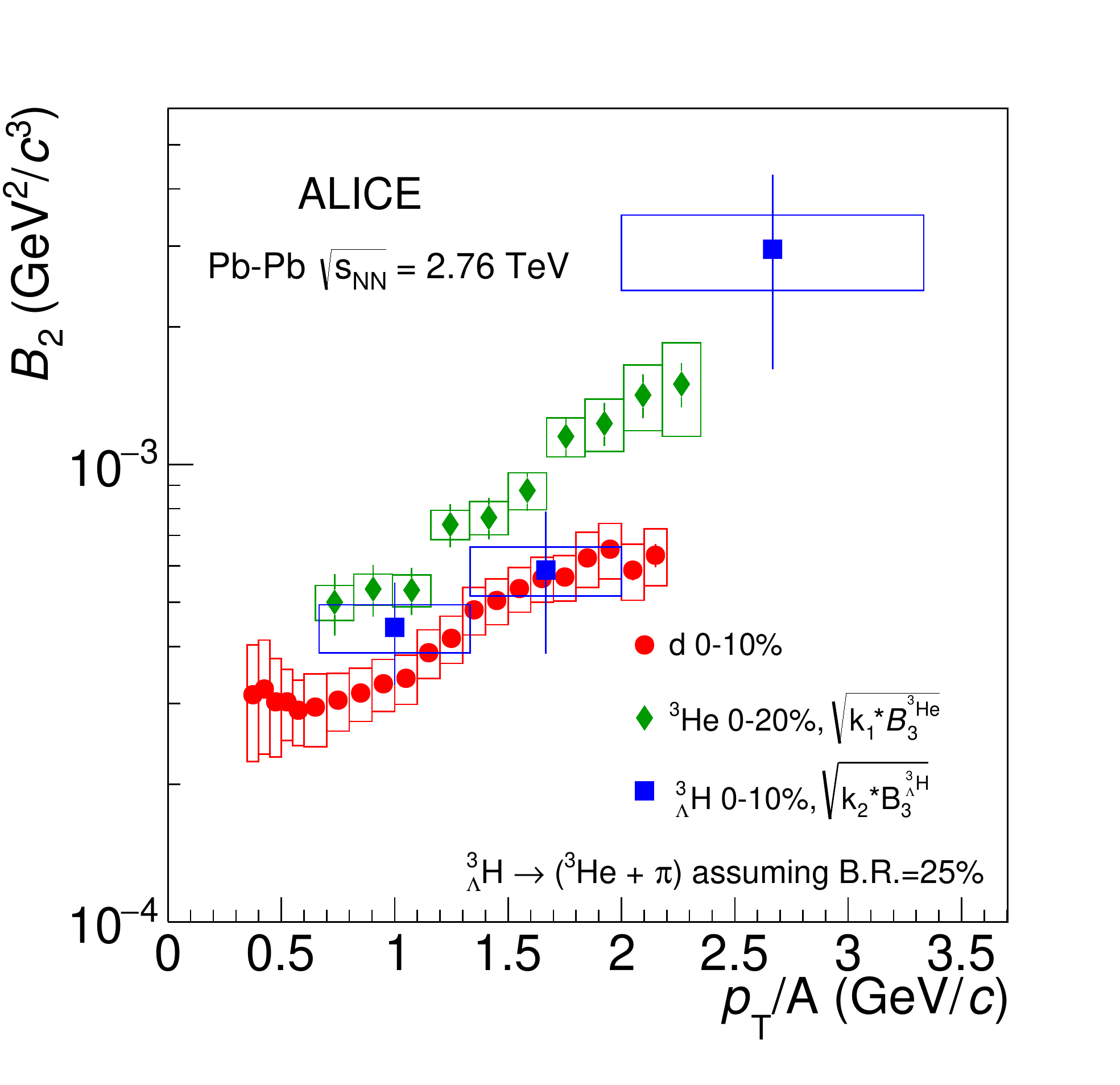}}
\end{minipage} & 
\begin{minipage}{.5\textwidth}
\centerline{\includegraphics[width=1\textwidth]{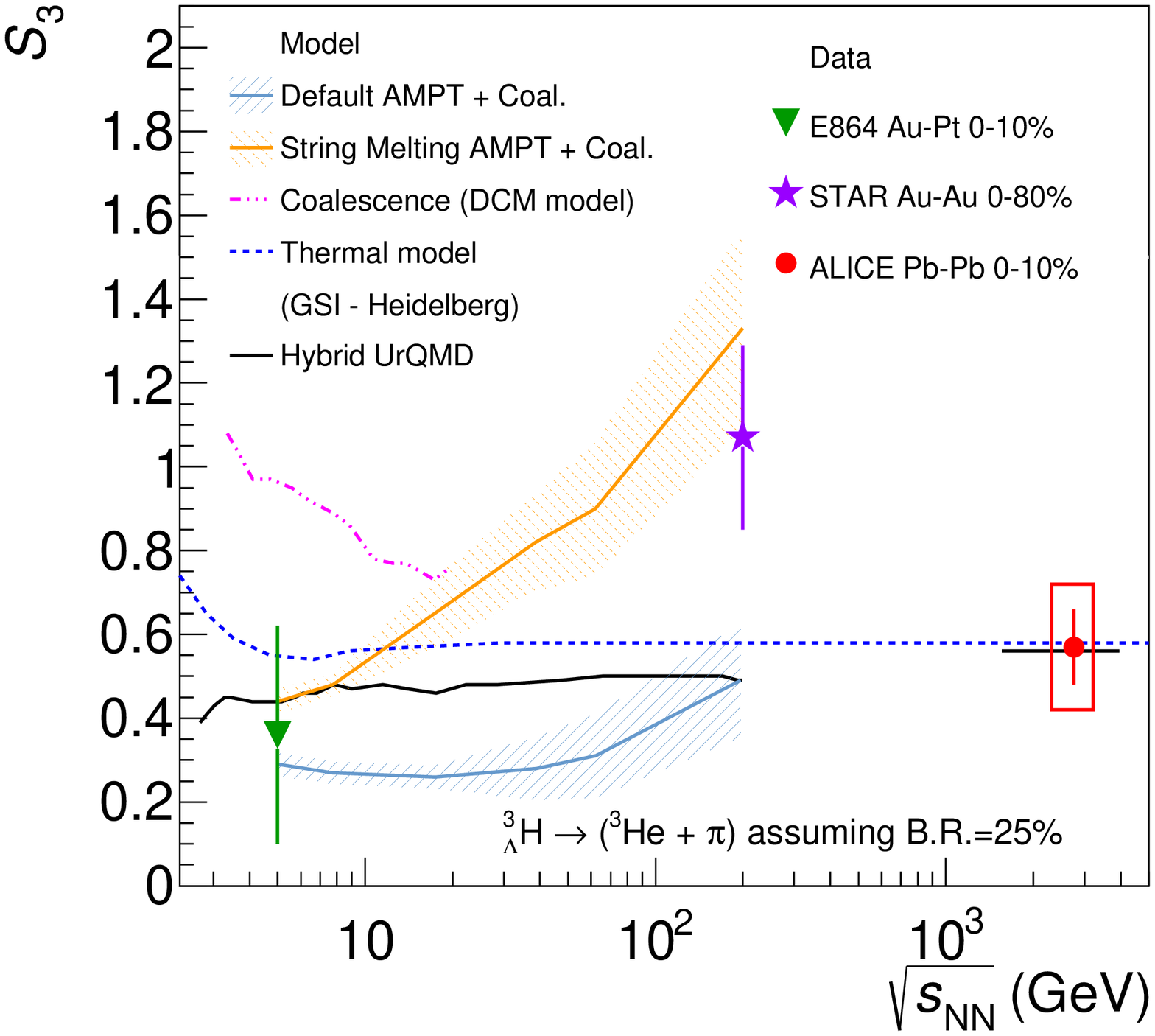}}
\end{minipage} 
\end{tabular}
\caption{Left: $B_2$ as a function of \pt\ /A for d (filled circles) \cite{Abbas:2014nuclei}, 
\he\ (empty circles) \cite{Abbas:2014nuclei},  and \hyp\ (filled squares). 
The $B_2^{(\mathrm{d},^{3}_{\Lambda}\mathrm{H})}$ and $B_2^{(\mathrm{d}, ^{3}\mathrm{He})}$
were evaluated as explained in the text. k$_1 = \frac{m_\mathrm{d}^2}{m_{^{3}\mathrm{He}}m_{\mathrm{p}}}$, 
and k$_2 = \frac{m_\mathrm{d}^2 m_{\Lambda}}{m_{\mathrm{p}}^2 m_{^{3}_{\Lambda}\mathrm H}}$. 
Right: $S_{3}$ ratio measured in this analysis compared with previous experimental results 
(E864 \cite{Armstrong:2002xh} and STAR \cite{Abelev:2010sci} (triangle and star, 
 respectively)) and different theoretical models as indicated in the legend.}
\label{fig:S3B3}
\end{figure}

In a simple coalescence model the $B_A$ parameter for all the  light nuclei should have the same behaviour. 
The coalescence parameter of deuteron ($B_2^{\mathrm{d}}$) and the coalescence 
parameters of \he\ and \hyp\ ($B_3^{^{3}{\rm{He}}}$ and  $B_3^{^{3}_{\Lambda}\mathrm{H}}$) can be directly compared 
deriving the $B_2^{^{3}\mathrm{He}}$ and the $B_2^{^{3}_{\Lambda}\mathrm{H}}$ using equation \ref{eq:eq4}, 
equation \ref{eq:eq5} and equation \ref{eq:eq6}. 
%The $B_2^{^{3}_{\Lambda}\mathrm{H}}$ parameter evaluated from 
%$B_3^{^{3}_{\Lambda}\mathrm{H}}$ is compared in Figure~\ref{fig:S3B3}  to that of 
%the deuteron, $B_2^\mathrm{d}$, and \he, $B_2^{^{3}\mathrm{He}}$. 
The comparison of the three coalescence parameters is shown in the left panel of Figure~\ref{fig:S3B3}.
The \hyp\ coalescence parameter is not flat as a function of \pt\ contrary to the prediction 
of the simple coalescence model \cite{Csernai:1986qf}, which does not take into account the characteristics of the emitting source. 
This is the same behaviour as observed for deuterons and \he\ nuclei \cite{Abbas:2014nuclei}.
At low \pt\ the $B_2$ values are compatible, suggesting that $p_0$ is similar for A = 2 and A = 3.\\
Using the measured \hyp\ yield the ratio 
$S_3 = $\hyp/(\he $\times \Lambda/\mathrm{p}$), also known as the strangeness population factor 
\cite{Zhang:2009ba}, was evaluated. 
%ORIGINAL
%This ratio was first suggested by the authors of \cite{Armstrong:2002xh} in the 
%expectation that dividing the strange to non-strange baryon yield should result 
%in a value near unity in a simple coalescence model. According to the authors 
%of \cite{Zhang:2009ba}, $S_3$ should be also a valuable tool to probe the nature 
%of the matter created in the collision, since it is sensitive to the local 
%baryon-strangeness correlation \cite{Koch:2005vg, Majumder:2006nq,Cheng:2008zh}.  
%In the thermal model approach the $S_3$ ratio does not depend on the chemical potential of particles 
%and was found to be almost energy independent \cite{Andronic:2011thermal, Steinheimer:2012tb}, 
%while in a dynamical coalescence picture it increases with decreasing beam energy 
%and is in general larger than the thermal model predictions \cite{Steinheimer:2012tb}. 
%This leads to the conclusion that the information on correlations of baryon number 
%and strangeness is lost in the thermal calculation because $S_3$ essentially depends only on 
%the temperature. On the other hand, in the microscopic treatment the correlation survives and
%$S_3$ provides information about the correlation between baryon and strangeness content.
%Finally, it has been argued \cite{Koch:2005vg,Zhang:2009ba} that a value of $S_3$ 
%close to unity indicates that the phase-space populations for strange and light quarks are 
%similar and would support the formation of high-temperature matter of deconfined quarks. 
This ratio was first suggested by the authors of \cite{Armstrong:2002xh} in the 
expectation that dividing the strange to non-strange baryon yield should result 
in a value near unity in a simple coalescence model. According to the authors 
of \cite{Zhang:2009ba}, $S_3$ should be also a valuable tool to probe the nature 
of the matter created in the collision, since it is sensitive to the local 
baryon-strangeness correlation \cite{Koch:2005vg, Majumder:2006nq,Cheng:2008zh}:
a value of $S_3$ close to unity would indicate that the phase-space populations for strange 
and light quarks are similar and would support the formation of high-temperature matter 
of deconfined quarks. 
In the thermal model approach the $S_3$ ratio does not depend on the chemical potential of particles 
and was found to be almost energy independent \cite{Andronic:2011thermal, Steinheimer:2012tb}, 
while in a dynamical coalescence picture it increases with decreasing beam energy 
and is in general larger than the thermal model predictions \cite{Steinheimer:2012tb}. 
This leads to the conclusion that the information on correlations of baryon number 
and strangeness is lost in the thermal calculation because $S_3$ essentially depends only on 
the temperature. 
The $\Lambda/\rm p$ ratio used in the present analysis was taken from \cite{Abelev:2013vea} 
and \cite{Abelev:2013lk0}.
The $S_3$ values obtained  for particles (anti-particles) are summarised in Table~\ref{table:s3} 
and the average of the two measurements is shown in the right panel of 
Figure~\ref{fig:S3B3}. 
These values were compared with different theoretical models  and to the results from 
experiments at BNL-AGS \cite{Armstrong:2002xh} and RHIC \cite{Abelev:2010sci}.\\
The models used for the comparison are the statistical hadronization model 
\cite{Andronic:2011thermal}, the hybrid UrQMD model \cite{Steinheimer:2012tb} and its extension at 
the LHC energy \cite{uqrmd:missing}, the DCM (Dubna Cascade Model) coalescence model 
(described in \cite{Steinheimer:2012tb}) 
and two versions -- default and string melting -- of the AMPT (A Multi-Phase Transport Model for 
Relativistic Heavy Ion Collisions) \cite{Lin:2004en} plus coalescence described 
in \cite{Zhang:2009ba}.
The present result at $\sqrt{s_{\rm{NN}}}$~=~2.76~TeV 
is comparable to that measured at E864 experiment \cite{Armstrong:2002xh} at $\sqrt{s_{\rm{NN}}}~\sim$~5~GeV, 
while it does not confirm the rising behaviour shown by STAR \cite{Abelev:2010sci} and by 
the AMPT with string melting plus coalescence model \cite{Zhang:2009ba}. 
This result is consistent with the thermal model approach, which predicts 
a constant S$_3$ value from \s\ above a few GeV.

\begin{table}[!h]
\def\arraystretch{1.5} 
\begin{center}
\begin{tabular}{ c | c | c }
\hline
\hline Centrality               &  $ \frac{^{3}_{\Lambda}\mathrm H}{^{3}\rm{He}} \times \frac{\rm p}{\Lambda} $        &   $\frac{^{3}_{\Lambda}\overline{\mathrm H}}{^{3}\rm{\overline {He}}} \times \frac{\rm \overline p}{\overline \Lambda} $ \\
\hline  0--10\%  &  $0.60 \pm 0.13 (\mathrm{stat.})\pm 0.21 (\mathrm{syst.})$  &  $0.54 \pm 0.13(\mathrm{stat.}) \pm 0.19 (\mathrm{syst.})$\\
\hline 
\end{tabular} 
\caption {$S_3$ for matter and anti-matter. To compute the ratio a B.R. of 25\% was assumed 
for the \hyp~$\rightarrow$~\he+~\pio\ decay.} 
\label{table:s3}
\end{center}
\end{table}

\section{Conclusions}
\label{sec:conclusions}
Measurements of \hyp\ and \antihyp\ in \PbPb\ collisions at \s~=~2.76~TeV were presented in this letter. 
The \hyp\ lifetime was measured and was found to agree with previous measurements 
within uncertainties. 
The measured value was included in the computation of the world average of the \hyp\ lifetime.
Transverse momentum yields at mid-rapidity for central (0--10\%) \PbPb\ 
collisions at \s~=~2.76 TeV were measured in three \pt\ intervals. 
The yields of particles and anti-particles were measured  in two centrality 
classes \mbox{(0--10\% and 10--50\%)} and compared with different 
theoretical models. The ratio \antihyp/\hyp\ is consistent with unity, as 
expected at the LHC energy. 
The measured yields indicate that hypernuclei in high-energy heavy-ion 
collisions are produced within an equilibrated thermal environment in which the 
temperature is the same as for the other particles produced at the LHC. 
The \hyp/\he\ (\antihyp/\antihe) ratio was also measured and compared with different 
theoretical models and results from the STAR experiment. 
STAR results are higher than ALICE results, but compatible within uncertainties.
The \hyp\ coalescence parameter was also evaluated. Its value increases with \pt, and 
within the uncertainties, is consistent with those extracted for deuteron and \he\ 
nuclei \cite{Abbas:2014nuclei}. 
The ratio $S_3 = $\hyp/(\he $\times \Lambda/\mathrm{p}$) was evaluated and compared with 
different theoretical models and measurements from previous experiments. 
The value of $S_3$ suggests that the production of nuclei and hypernuclei at the LHC  
can be described with a thermodynamic approach, and is similar to the one calculated 
by the Hybrid UrQMD model \cite{uqrmd:missing}.
No conclusions can be drawn about the AMPT + coalescence model \cite{Zhang:2009ba}, 
since no prediction of dynamical coalescence  models is available at the LHC energy. 
The measured $S_3$ value excludes the rising trend in AMPT seen up to RHIC energies 
extends to LHC energies. 
The $S_3$ measured at AGS, RHIC and LHC are compatible within uncertainty with a value which is 
independent of the centre of mass energy of the collision. 
 
%%%%%%%%% acknowledgements
\newenvironment{acknowledgement}{\relax}{\relax}
\begin{acknowledgement}
\section*{Acknowledgements}
% $Id: acknowledgements.tex 2098 2015-04-24 15:54:16Z loizides $
% Version: Jan 2015

The ALICE Collaboration would like to thank all its engineers and technicians for their invaluable contributions to the construction of the experiment and the CERN accelerator teams for the outstanding performance of the LHC complex.
The ALICE Collaboration gratefully acknowledges the resources and support provided by all Grid centres and the Worldwide LHC Computing Grid (WLCG) collaboration.
The ALICE Collaboration acknowledges the following funding agencies for their support in building and
running the ALICE detector:
State Committee of Science,  World Federation of Scientists (WFS)
and Swiss Fonds Kidagan, Armenia,
Conselho Nacional de Desenvolvimento Cient\'{\i}fico e Tecnol\'{o}gico (CNPq), Financiadora de Estudos e Projetos (FINEP),
Funda\c{c}\~{a}o de Amparo \`{a} Pesquisa do Estado de S\~{a}o Paulo (FAPESP);
National Natural Science Foundation of China (NSFC), the Chinese Ministry of Education (CMOE)
and the Ministry of Science and Technology of China (MSTC);
Ministry of Education and Youth of the Czech Republic;
Danish Natural Science Research Council, the Carlsberg Foundation and the Danish National Research Foundation;
The European Research Council under the European Community's Seventh Framework Programme;
Helsinki Institute of Physics and the Academy of Finland;
French CNRS-IN2P3, the `Region Pays de Loire', `Region Alsace', `Region Auvergne' and CEA, France;
German Bundesministerium fur Bildung, Wissenschaft, Forschung und Technologie (BMBF) and the Helmholtz Association;
General Secretariat for Research and Technology, Ministry of
Development, Greece;
Hungarian Orszagos Tudomanyos Kutatasi Alappgrammok (OTKA) and National Office for Research and Technology (NKTH);
Department of Atomic Energy and Department of Science and Technology of the Government of India;
Istituto Nazionale di Fisica Nucleare (INFN) and Centro Fermi -
Museo Storico della Fisica e Centro Studi e Ricerche "Enrico
Fermi", Italy;
MEXT Grant-in-Aid for Specially Promoted Research, Ja\-pan;
Joint Institute for Nuclear Research, Dubna;
National Research Foundation of Korea (NRF);
Consejo Nacional de Cienca y Tecnologia (CONACYT), Direccion General de Asuntos del Personal Academico(DGAPA), M\'{e}xico, Amerique Latine Formation academique - European Commission~(ALFA-EC) and the EPLANET Program~(European Particle Physics Latin American Network);
Stichting voor Fundamenteel Onderzoek der Materie (FOM) and the Nederlandse Organisatie voor Wetenschappelijk Onderzoek (NWO), Netherlands;
Research Council of Norway (NFR);
National Science Centre, Poland;
Ministry of National Education/Institute for Atomic Physics and National Council of Scientific Research in Higher Education~(CNCSI-UEFISCDI), Romania;
Ministry of Education and Science of Russian Federation, Russian
Academy of Sciences, Russian Federal Agency of Atomic Energy,
Russian Federal Agency for Science and Innovations and The Russian
Foundation for Basic Research;
Ministry of Education of Slovakia;
Department of Science and Technology, South Africa;
Centro de Investigaciones Energeticas, Medioambientales y Tecnologicas (CIEMAT), E-Infrastructure shared between Europe and Latin America (EELA), Ministerio de Econom\'{i}a y Competitividad (MINECO) of Spain, Xunta de Galicia (Conseller\'{\i}a de Educaci\'{o}n),
Centro de Aplicaciones Tecnológicas y Desarrollo Nuclear (CEA\-DEN), Cubaenerg\'{\i}a, Cuba, and IAEA (International Atomic Energy Agency);
Swedish Research Council (VR) and Knut $\&$ Alice Wallenberg
Foundation (KAW);
Ukraine Ministry of Education and Science;
United Kingdom Science and Technology Facilities Council (STFC);
The United States Department of Energy, the United States National
Science Foundation, the State of Texas, and the State of Ohio;
Ministry of Science, Education and Sports of Croatia and  Unity through Knowledge Fund, Croatia.
Council of Scientific and Industrial Research (CSIR), New Delhi, India
    %%%%%%% get the lates version before submitting
\end{acknowledgement}

\bibliographystyle{utphys}
\bibliography{bibliog.bib}

\newpage

%%%%%%%%% appendix with author list
\appendix
\section{The ALICE Collaboration}
\label{app:collab}

% Collaboration: CERN-LHC-ALICE
% Generation Date is 2015/Apr/15

% How to use:
%%%%%%%%% appendix with author list
%\appendix
%\section{The ALICE Collaboration}
%\label{app:collab}
%\input{authors-list.tex}  %%%%%%% get the latest version before submitting

\begingroup
\small
\begin{flushleft}
J.~Adam\Irefn{org40}\And
D.~Adamov\'{a}\Irefn{org83}\And
M.M.~Aggarwal\Irefn{org87}\And
G.~Aglieri Rinella\Irefn{org36}\And
M.~Agnello\Irefn{org111}\And
N.~Agrawal\Irefn{org48}\And
Z.~Ahammed\Irefn{org132}\And
S.U.~Ahn\Irefn{org68}\And
I.~Aimo\Irefn{org94}\textsuperscript{,}\Irefn{org111}\And
S.~Aiola\Irefn{org137}
% ...
\renewcommand\labelenumi{\textsuperscript{\theenumi}~}

\section*{Affiliation notes}
\renewcommand\theenumi{\roman{enumi}}
\begin{Authlist}
\item \Adef{0}Deceased
\item \Adef{idp5907008}{Also at: University of Kansas, Lawrence, Kansas, United States}
\end{Authlist}

\section*{Collaboration Institutes}
\renewcommand\theenumi{\arabic{enumi}~}
\begin{Authlist}

\item \Idef{org1}A.I. Alikhanyan National Science Laboratory (Yerevan Physics Institute) Foundation, Yerevan, Armenia
\item \Idef{org2}Benem\'{e}rita Universidad Aut\'{o}noma de Puebla, Puebla, Mexico
\item \Idef{org3}Bogolyubov Institute for Theoretical Physics, Kiev, Ukraine
\item \Idef{org4}Bose Institute, Department of Physics and Centre for Astroparticle Physics and Space Science (CAPSS), Kolkata, India
\item \Idef{org5}Budker Institute for Nuclear Physics, Novosibirsk, Russia
\item \Idef{org6}California Polytechnic State University, San Luis Obispo, California, United States
\item \Idef{org7}Central China Normal University, Wuhan, China
\item \Idef{org8}Centre de Calcul de l'IN2P3, Villeurbanne, France
\item \Idef{org9}Centro de Aplicaciones Tecnol\'{o}gicas y Desarrollo Nuclear (CEADEN), Havana, Cuba
\item \Idef{org10}Centro de Investigaciones Energ\'{e}ticas Medioambientales y Tecnol\'{o}gicas (CIEMAT), Madrid, Spain
\item \Idef{org11}Centro de Investigaci\'{o}n y de Estudios Avanzados (CINVESTAV), Mexico City and M\'{e}rida, Mexico
\item \Idef{org12}Centro Fermi - Museo Storico della Fisica e Centro Studi e Ricerche ``Enrico Fermi'', Rome, Italy
\item \Idef{org13}Chicago State University, Chicago, Illinois, USA
\item \Idef{org14}China Institute of Atomic Energy, Beijing, China
\item \Idef{org15}Commissariat \`{a} l'Energie Atomique, IRFU, Saclay, France
\item \Idef{org16}COMSATS Institute of Information Technology (CIIT), Islamabad, Pakistan
\item \Idef{org17}Departamento de F\'{\i}sica de Part\'{\i}culas and IGFAE, Universidad de Santiago de Compostela, Santiago de Compostela, Spain
\item \Idef{org18}Department of Physics and Technology, University of Bergen, Bergen, Norway
\item \Idef{org19}Department of Physics, Aligarh Muslim University, Aligarh, India
\item \Idef{org20}Department of Physics, Ohio State University, Columbus, Ohio, United States
\item \Idef{org21}Department of Physics, Sejong University, Seoul, South Korea
\item \Idef{org22}Department of Physics, University of Oslo, Oslo, Norway
\item \Idef{org23}Dipartimento di Elettrotecnica ed Elettronica del Politecnico, Bari, Italy
\item \Idef{org24}Dipartimento di Fisica dell'Universit\`{a} 'La Sapienza' and Sezione INFN Rome, Italy
\item \Idef{org25}Dipartimento di Fisica dell'Universit\`{a} and Sezione INFN, Cagliari, Italy
\item \Idef{org26}Dipartimento di Fisica dell'Universit\`{a} and Sezione INFN, Trieste, Italy
\item \Idef{org27}Dipartimento di Fisica dell'Universit\`{a} and Sezione INFN, Turin, Italy
\item \Idef{org28}Dipartimento di Fisica e Astronomia dell'Universit\`{a} and Sezione INFN, Bologna, Italy
\item \Idef{org29}Dipartimento di Fisica e Astronomia dell'Universit\`{a} and Sezione INFN, Catania, Italy
\item \Idef{org30}Dipartimento di Fisica e Astronomia dell'Universit\`{a} and Sezione INFN, Padova, Italy
\item \Idef{org31}Dipartimento di Fisica `E.R.~Caianiello' dell'Universit\`{a} and Gruppo Collegato INFN, Salerno, Italy
\item \Idef{org32}Dipartimento di Scienze e Innovazione Tecnologica dell'Universit\`{a} del  Piemonte Orientale and Gruppo Collegato INFN, Alessandria, Italy
\item \Idef{org33}Dipartimento Interateneo di Fisica `M.~Merlin' and Sezione INFN, Bari, Italy
\item \Idef{org34}Division of Experimental High Energy Physics, University of Lund, Lund, Sweden
\item \Idef{org35}Eberhard Karls Universit\"{a}t T\"{u}bingen, T\"{u}bingen, Germany
\item \Idef{org36}European Organization for Nuclear Research (CERN), Geneva, Switzerland
\item \Idef{org37}Excellence Cluster Universe, Technische Universit\"{a}t M\"{u}nchen, Munich, Germany
\item \Idef{org38}Faculty of Engineering, Bergen University College, Bergen, Norway
\item \Idef{org39}Faculty of Mathematics, Physics and Informatics, Comenius University, Bratislava, Slovakia
\item \Idef{org40}Faculty of Nuclear Sciences and Physical Engineering, Czech Technical University in Prague, Prague, Czech Republic
\item \Idef{org41}Faculty of Science, P.J.~\v{S}af\'{a}rik University, Ko\v{s}ice, Slovakia
\item \Idef{org42}Faculty of Technology, Buskerud and Vestfold University College, Vestfold, Norway
\item \Idef{org43}Frankfurt Institute for Advanced Studies, Johann Wolfgang Goethe-Universit\"{a}t Frankfurt, Frankfurt, Germany
\item \Idef{org44}Gangneung-Wonju National University, Gangneung, South Korea
\item \Idef{org45}Gauhati University, Department of Physics, Guwahati, India
\item \Idef{org46}Helsinki Institute of Physics (HIP), Helsinki, Finland
\item \Idef{org47}Hiroshima University, Hiroshima, Japan
\item \Idef{org48}Indian Institute of Technology Bombay (IIT), Mumbai, India
\item \Idef{org49}Indian Institute of Technology Indore, Indore (IITI), India
\item \Idef{org50}Inha University, Incheon, South Korea
\item \Idef{org51}Institut de Physique Nucl\'eaire d'Orsay (IPNO), Universit\'e Paris-Sud, CNRS-IN2P3, Orsay, France
\item \Idef{org52}Institut f\"{u}r Informatik, Johann Wolfgang Goethe-Universit\"{a}t Frankfurt, Frankfurt, Germany
\item \Idef{org53}Institut f\"{u}r Kernphysik, Johann Wolfgang Goethe-Universit\"{a}t Frankfurt, Frankfurt, Germany
\item \Idef{org54}Institut f\"{u}r Kernphysik, Westf\"{a}lische Wilhelms-Universit\"{a}t M\"{u}nster, M\"{u}nster, Germany
\item \Idef{org55}Institut Pluridisciplinaire Hubert Curien (IPHC), Universit\'{e} de Strasbourg, CNRS-IN2P3, Strasbourg, France
\item \Idef{org56}Institute for Nuclear Research, Academy of Sciences, Moscow, Russia
\item \Idef{org57}Institute for Subatomic Physics of Utrecht University, Utrecht, Netherlands
\item \Idef{org58}Institute for Theoretical and Experimental Physics, Moscow, Russia
\item \Idef{org59}Institute of Experimental Physics, Slovak Academy of Sciences, Ko\v{s}ice, Slovakia
\item \Idef{org60}Institute of Physics, Academy of Sciences of the Czech Republic, Prague, Czech Republic
\item \Idef{org61}Institute of Physics, Bhubaneswar, India
\item \Idef{org62}Institute of Space Science (ISS), Bucharest, Romania
\item \Idef{org63}Instituto de Ciencias Nucleares, Universidad Nacional Aut\'{o}noma de M\'{e}xico, Mexico City, Mexico
\item \Idef{org64}Instituto de F\'{\i}sica, Universidad Nacional Aut\'{o}noma de M\'{e}xico, Mexico City, Mexico
\item \Idef{org65}iThemba LABS, National Research Foundation, Somerset West, South Africa
\item \Idef{org66}Joint Institute for Nuclear Research (JINR), Dubna, Russia
\item \Idef{org67}Konkuk University, Seoul, South Korea
\item \Idef{org68}Korea Institute of Science and Technology Information, Daejeon, South Korea
\item \Idef{org69}KTO Karatay University, Konya, Turkey
\item \Idef{org70}Laboratoire de Physique Corpusculaire (LPC), Clermont Universit\'{e}, Universit\'{e} Blaise Pascal, CNRS--IN2P3, Clermont-Ferrand, France
\item \Idef{org71}Laboratoire de Physique Subatomique et de Cosmologie, Universit\'{e} Grenoble-Alpes, CNRS-IN2P3, Grenoble, France
\item \Idef{org72}Laboratori Nazionali di Frascati, INFN, Frascati, Italy
\item \Idef{org73}Laboratori Nazionali di Legnaro, INFN, Legnaro, Italy
\item \Idef{org74}Lawrence Berkeley National Laboratory, Berkeley, California, United States
\item \Idef{org75}Lawrence Livermore National Laboratory, Livermore, California, United States
\item \Idef{org76}Moscow Engineering Physics Institute, Moscow, Russia
\item \Idef{org77}National Centre for Nuclear Studies, Warsaw, Poland
\item \Idef{org78}National Institute for Physics and Nuclear Engineering, Bucharest, Romania
\item \Idef{org79}National Institute of Science Education and Research, Bhubaneswar, India
\item \Idef{org80}Niels Bohr Institute, University of Copenhagen, Copenhagen, Denmark
\item \Idef{org81}Nikhef, National Institute for Subatomic Physics, Amsterdam, Netherlands
\item \Idef{org82}Nuclear Physics Group, STFC Daresbury Laboratory, Daresbury, United Kingdom
\item \Idef{org83}Nuclear Physics Institute, Academy of Sciences of the Czech Republic, \v{R}e\v{z} u Prahy, Czech Republic
\item \Idef{org84}Oak Ridge National Laboratory, Oak Ridge, Tennessee, United States
\item \Idef{org85}Petersburg Nuclear Physics Institute, Gatchina, Russia
\item \Idef{org86}Physics Department, Creighton University, Omaha, Nebraska, United States
\item \Idef{org87}Physics Department, Panjab University, Chandigarh, India
\item \Idef{org88}Physics Department, University of Athens, Athens, Greece
\item \Idef{org89}Physics Department, University of Cape Town, Cape Town, South Africa
\item \Idef{org90}Physics Department, University of Jammu, Jammu, India
\item \Idef{org91}Physics Department, University of Rajasthan, Jaipur, India
\item \Idef{org92}Physik Department, Technische Universit\"{a}t M\"{u}nchen, Munich, Germany
\item \Idef{org93}Physikalisches Institut, Ruprecht-Karls-Universit\"{a}t Heidelberg, Heidelberg, Germany
\item \Idef{org94}Politecnico di Torino, Turin, Italy
\item \Idef{org95}Purdue University, West Lafayette, Indiana, United States
\item \Idef{org96}Pusan National University, Pusan, South Korea
\item \Idef{org97}Research Division and ExtreMe Matter Institute EMMI, GSI Helmholtzzentrum f\"ur Schwerionenforschung, Darmstadt, Germany
\item \Idef{org98}Rudjer Bo\v{s}kovi\'{c} Institute, Zagreb, Croatia
\item \Idef{org99}Russian Federal Nuclear Center (VNIIEF), Sarov, Russia
\item \Idef{org100}Russian Research Centre Kurchatov Institute, Moscow, Russia
\item \Idef{org101}Saha Institute of Nuclear Physics, Kolkata, India
\item \Idef{org102}School of Physics and Astronomy, University of Birmingham, Birmingham, United Kingdom
\item \Idef{org103}Secci\'{o}n F\'{\i}sica, Departamento de Ciencias, Pontificia Universidad Cat\'{o}lica del Per\'{u}, Lima, Peru
\item \Idef{org104}Sezione INFN, Bari, Italy
\item \Idef{org105}Sezione INFN, Bologna, Italy
\item \Idef{org106}Sezione INFN, Cagliari, Italy
\item \Idef{org107}Sezione INFN, Catania, Italy
\item \Idef{org108}Sezione INFN, Padova, Italy
\item \Idef{org109}Sezione INFN, Rome, Italy
\item \Idef{org110}Sezione INFN, Trieste, Italy
\item \Idef{org111}Sezione INFN, Turin, Italy
\item \Idef{org112}SSC IHEP of NRC Kurchatov institute, Protvino, Russia
\item \Idef{org113}SUBATECH, Ecole des Mines de Nantes, Universit\'{e} de Nantes, CNRS-IN2P3, Nantes, France
\item \Idef{org114}Suranaree University of Technology, Nakhon Ratchasima, Thailand
\item \Idef{org115}Technical University of Ko\v{s}ice, Ko\v{s}ice, Slovakia
\item \Idef{org116}Technical University of Split FESB, Split, Croatia
\item \Idef{org117}The Henryk Niewodniczanski Institute of Nuclear Physics, Polish Academy of Sciences, Cracow, Poland
\item \Idef{org118}The University of Texas at Austin, Physics Department, Austin, Texas, USA
\item \Idef{org119}Universidad Aut\'{o}noma de Sinaloa, Culiac\'{a}n, Mexico
\item \Idef{org120}Universidade de S\~{a}o Paulo (USP), S\~{a}o Paulo, Brazil
\item \Idef{org121}Universidade Estadual de Campinas (UNICAMP), Campinas, Brazil
\item \Idef{org122}University of Houston, Houston, Texas, United States
\item \Idef{org123}University of Jyv\"{a}skyl\"{a}, Jyv\"{a}skyl\"{a}, Finland
\item \Idef{org124}University of Liverpool, Liverpool, United Kingdom
\item \Idef{org125}University of Tennessee, Knoxville, Tennessee, United States
\item \Idef{org126}University of the Witwatersrand, Johannesburg, South Africa
\item \Idef{org127}University of Tokyo, Tokyo, Japan
\item \Idef{org128}University of Tsukuba, Tsukuba, Japan
\item \Idef{org129}University of Zagreb, Zagreb, Croatia
\item \Idef{org130}Universit\'{e} de Lyon, Universit\'{e} Lyon 1, CNRS/IN2P3, IPN-Lyon, Villeurbanne, France
\item \Idef{org131}V.~Fock Institute for Physics, St. Petersburg State University, St. Petersburg, Russia
\item \Idef{org132}Variable Energy Cyclotron Centre, Kolkata, India
\item \Idef{org133}Vin\v{c}a Institute of Nuclear Sciences, Belgrade, Serbia
\item \Idef{org134}Warsaw University of Technology, Warsaw, Poland
\item \Idef{org135}Wayne State University, Detroit, Michigan, United States
\item \Idef{org136}Wigner Research Centre for Physics, Hungarian Academy of Sciences, Budapest, Hungary
\item \Idef{org137}Yale University, New Haven, Connecticut, United States
\item \Idef{org138}Yonsei University, Seoul, South Korea
\item \Idef{org139}Zentrum f\"{u}r Technologietransfer und Telekommunikation (ZTT), Fachhochschule Worms, Worms, Germany
\end{Authlist}
\endgroup

  %%%%%%% get the latest version before submitting
\end{document}